# Search Space Engine Optimize Search Using FCC_STF Algorithm in Fuzzy Co-Clustering

Monika Rani, Anubha Parashar, Jyoti Chaturvedi and Anu Malviya

*Abstract*— Fuzzy co-clustering can be improved if we handle two main problem first is outlier and second curse of dimensionality .outlier problem can be reduce by implementing page replacement algorithm like FIFO, LRU or priority algorithm in a set of frame of web pages efficiently through a search engine. The web page which has zero priority (outlier) can be represented in separate slot of frame. Whereas curse of dimensionality problem can be improved by implementing FCC_STF algorithm for web pages obtain by search engine that reduce the outlier problem first. The algorithm FCCM and FUZZY CO-DOK are compared with FCC_STF algorithm with merit and demerits on the bases of different fuzzifier used. FCC_STF algorithm in which fuzzifier fused into one entity who have shown high performance by experiment result of values (A1,B1,Vcj,A2,B2) seem to less sensitive to local maxima and obtain optimization search space in 2-D for web pages by plotting graph between J(fcc_stf ) and Vcj.

## I. INTRODUCTION

Fuzzy clustering method is offered to construct clusters with uncertain to boundaries and allow that one object to belong to one or more cluster with some membership degree. Fuzzy co-clustering (bi-clustering) is a technique to simultaneously cluster data (web document) and feature(words inside the document).Fuzzy co-clustering have some advantages like dimensionality reduction, interpretable document cluster, improvement in accuracy due to local modal of clustering.

In this thesis solution for outlier and curse of dimensionality is provided. To reduce outlier fuzzy membership of each web document is found, so each document belong to particular cluster base on the priority of page assign in frame of five with additional to sixth frame of zero web page priority. And second problem is cure of dimensionality-fuzzy co-clustering

*Monika Rani*, Department of Computer Science & Engineering ,YMCA University of science and Technology, Faridabad, India, 08860130090,

with Ruspini's condition (FCR) and later FCCM [4] & fuzzy codok are added to this family with advance FCC_STF. In FCC_STF(Fuzzy co-clustering with single term fuzzifier) instead of two entities (like in fuzzy codok), here fuzzifier is fused into one entity.

## II. OUTLIER PROBLEMS

Data (web document) object that do not comply with the general behavior or model (cluster) of the data. These data object are outlier. Most data mining method discard outliers as noise or exceptions. However, in some application such as fraud detection, the rare event can be more interesting than the more regularly occurring ones. The analysis of outlier data is referred to as outlier mining.

Outliers may be detected using statistical test that assume a distribution or probability model for the data or using distance measure where object that are a substantial distance from any other cluster are considered outliers. Rather than using statistical or distance measures, deviation-based methods identify method identify outlier by examining differences in the main characteristics of object in a group.

To reduce outlier fuzzy membership of each web document is found, so each document belong to particular cluster. Thus searching of web document becomes efficient.

Here five frames are considered where we can use page replacement algorithm like.

A. FIFO (Fist in First Out)

B. LRU (Least Recently Use)

But as per preference here, PRIORITY (number of visit) basis is chooses mean on the basis of the priority page is set in the form of size 5 which the search engine will give the output for the high priority pages as given in the description (description name) by the description link.

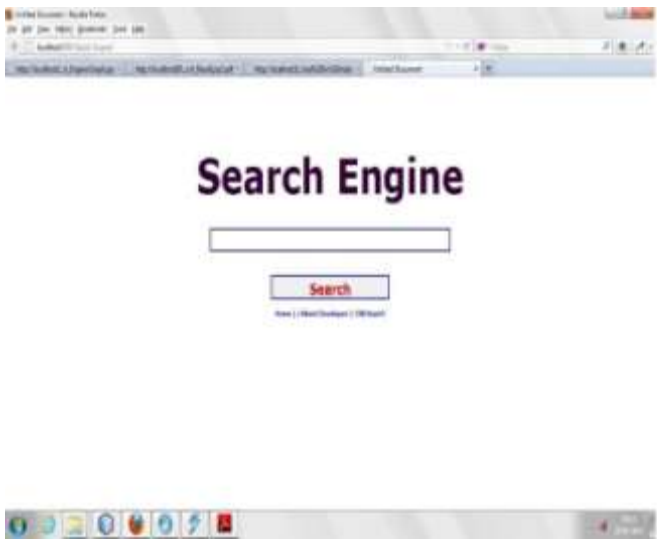

. Fig.1 Search Engine without query

*A. Search Engine*

This search engine works in this manner as:

• Link Name: The source name is describe here from which data has to be taken means source for web pages .These web pages refer when the link description define the keywords and these keywords are search in source (data base ) for particular web page.

• Link Description: Here we mention the statement (set of keyword) for which specific data is represented in web pages.

• Keyword: Keyword mention help in finding required web page and the key word separate by comma define particular key word for particular web page. Basically following step take place :

1-obtaine set of keyword when enter in search engine.

2-tokenize the text in the search engine and for particular keyword find appropriate web page from database.

3- break the keyword (remove stop word ) , means if there are three key word then in search engine then , loop for k=3 , k=2 , k=1 is search bases for webpage in data base.

4- Do linguistic pre-processing a list of normalized tokens, which are present in web page by using key word as search bases of web pages.

5 –finally obtain the link for the web page in the five frames with high priority and one additional to these five frames6[th] with zero priority web pages.

*B. Example*

Link Name:- www.education.ac.in, education site is used in the link name with limit of varchar(45) to enter the character.

Link descriptor:- define the content show on the link or the matter present on the link its limit is define as varchar(450) to enter the character, here "we deal in all type of education" is given as link descriptor.

Keyword:– define the key word on which particular web page is found , the word limit of key word varchar (400) , by the key word match web page is brought from web data base.

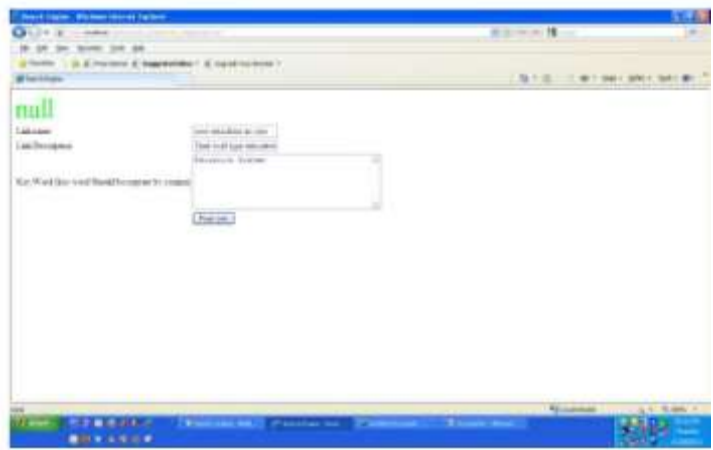

Fig.2 Link Registration

Once link name, descriptor and key word is mention then it is post. Below is view after post link.

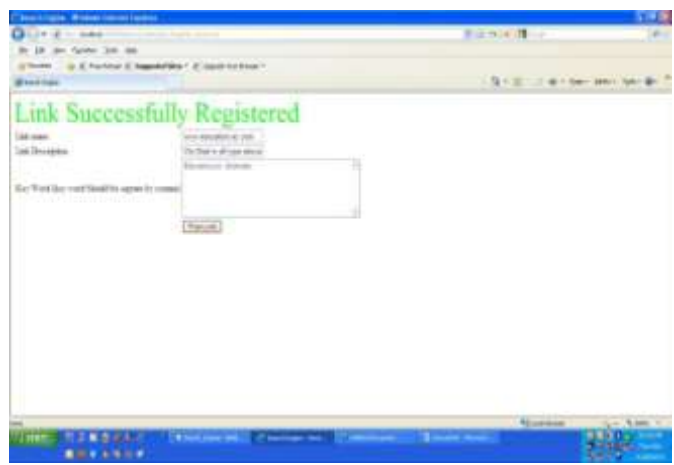

Fig.3 Link Successfully Registered

### III. EXPERIMENTAL RESULT

The search engine made to search on the bases of keyword, example –"education system parameter".

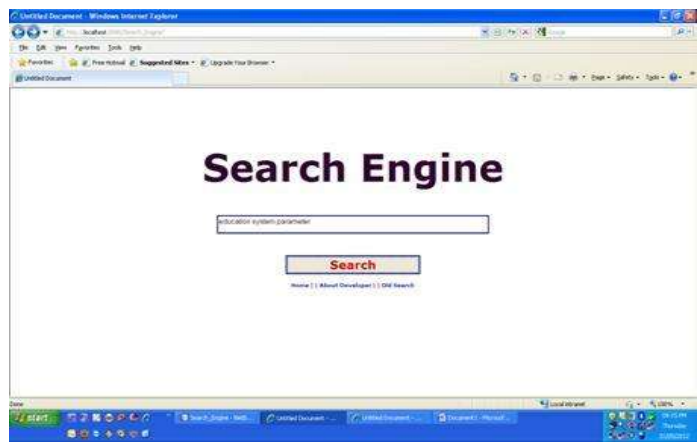

Fig.4 Search Engine with query

The main problem of fuzzy co-cluster is that the outlier are more in number, means there are web pages which lie outside the clusters.

Thus, this search engine reduces the number of outlier web pages by applying priority bases algorithm in five frame and soon on for next five frame with high priority and addition to this here $6^{th}$ is use to represent the web page whose priority is "ZERO" by this we get 5 pages of high priority along with zero priority page, total six frame and in this approach the "ZERO" priority pages also found which was previously an outlier only .Thus this approach give chance to find out web page with zero priority .

- "ZERO" priority page is also important as its link descriptor might not define correctly and that page contain information which we require, just because of bad approach of searching we can loss that web page as outlier, thus this approach help to find the appropriate information as per requirement .Also improve the efficiency to search web document. Thus we obtain appropriate information as per requirement.

- After executing the query search engine show the total number of row is 18 and six frame in which five frame denote the web page of high priority and additional one show the frame for "zero" priority.

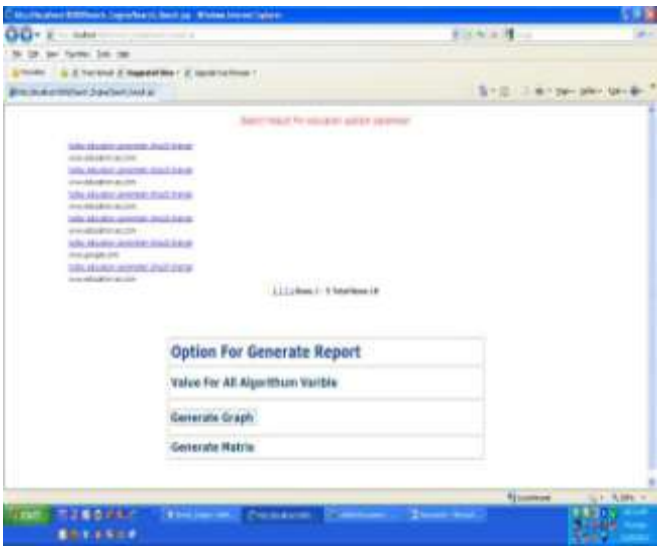

Fig.5 Frame for search engine on priority bases of web page

1. Set parameters C, Tu, Tv and E;
2. Randomly initialize Uci;
3. REPEAT
4. Update Vcj using (equation);
5. If any Vcj is negative
6. Perform clipping and renormalization on Vci;
7. Update Uci using (equation);
8. If any Uci is negative
9. Perform clipping an renormalizing on Uci;
10. Until max (|Uci(t+1)-Uci(t)|)<E.

Where C, N, K denotes the number of co-clusters, documents and words.

- Uci and Vcj denote membership of document and word respectively
- dij denote the correlation degree between document and words
- Tu and Tv are the degree of fuzziness parameters to adjust the level of fuzziness in the document and word clusters respectively.

The FCC_STF new algorithm is proposed as compare to FCCM and fuzzy codok, which discovering relevant information effectively. Here values of variable like A1, B1, Vcj, A2, and B2.

$$u_{ci} = \frac{2T\left(\sum_{j=1}^{K} v_{cj}^2 - 1\right) - \sum_{j=1}^{K} v_{cj} d_{ij} + (A_1/B_1)}{2T\left(\sum_{j=1}^{K} v_{cj}^2 - 1\right)}$$

$$v_{cj} = \frac{2T\left(\sum_{i=1}^{N} u_{ci}^2 - \sum_{i=1}^{N} u_{ci}\right) - \sum_{i=1}^{N} u_{ci} d_{ij} + (A_2/B_2)}{2T\left(1 - 2\sum_{i=1}^{N} u_{ci} + \sum_{i=1}^{N} u_{ci}^2\right)}$$

where:

- $A_1 = 1 - \sum_{d=1}^{C}\left(\frac{2T\left(\sum_{j=1}^{K} v_{cj}^2 - 1\right) - \sum_{j=1}^{K} v_{cj} d_{ij}}{2T\left(\sum_{j=1}^{K} v_{cj}^2 - 1\right)}\right)$

- $B_1 = \sum_{d=1}^{C}\left[1/\left\{2T\left(\sum_{j=1}^{K} v_{cj}^2 - 1\right)\right\}\right]$

- $A_2 = 1 - \sum_{q=1}^{K}\left(\frac{2T\left(\sum_{i=1}^{N} u_{ci}^2 - \sum_{i=1}^{N} u_{ci}\right) - \sum_{i=1}^{N} u_{ci} d_{ij}}{2T\left(1 - 2\sum_{i=1}^{N} u_{ci} + \sum_{i=1}^{N} u_{ci}^2\right)}\right)$

- $B_2 = K/\left\{2T\left(1 - 2\sum_{i=1}^{N} u_{ci} + \sum_{i=1}^{N} u_{ci}^2\right)\right\}$

Table.1 Value of variable in Fcc_Stf algorithm

IV. CURSE OF DIMENSIONALITY PROBLEM

The curse of dimensionality :refer so various phenomena that arise when analyzing and organizing high dimensional space (100-1000) that do not occur in low dimensional setting such as the physical space commonly modeled with just 3-D when considering problem in dynamic optimization .To solve curse of dimensionality Fcc_Stf algorithm[5] is used.

FCC-STF Algorithm

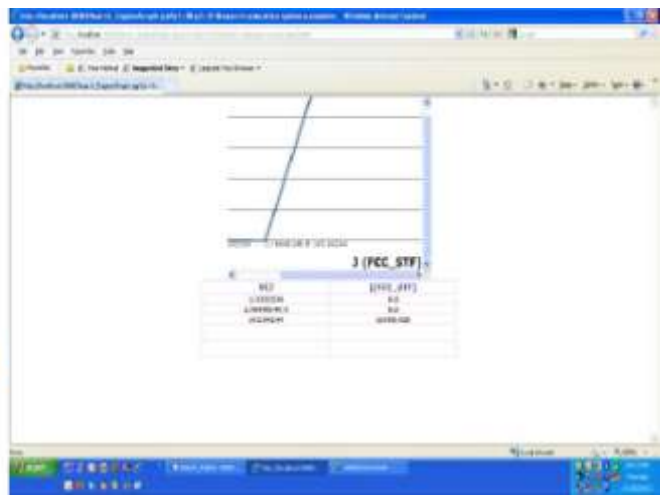

Optimize function J(fcc_stf) is defined for Fcc_Stf algorithm:

$$J_{FCC-STF} = \sum_{c=1}^{C}\sum_{i=1}^{N}\sum_{j=1}^{K} u_{ci} v_{cj} d_{ij} + T\sum_{c=1}^{C}\sum_{i=1}^{N}\sum_{j=1}^{K}\left[(u_{ci}+v_{cj})-u_{ci}v_{cj}\right]^2 + \sum_{i=1}^{N}\lambda_i\left(\sum_{c=1}^{C}u_{ci}-1\right)+\sum_{c=1}^{C}\gamma_c\left(\sum_{j=1}^{K}v_{cj}-1\right)$$

Table 2 Table of Vcj and J(fcc_stf) value

Optimize Search Space**:** The value of vcj and J(fcc_stf) is use find optimize space search graph in 2-D :-

Fig.6 Graph between Vcj and J(Fcc_Stf)

## V. CONCLUSION

**1.** The document get membership Fuzzy co-clustering provides efficient result in situation of vague and uncertainty. Fuzzy co-clustering could in many ways outperform standard fuzzy clustering when applied on a highly overlapping dataset.

2. To avoid the problem of always getting one fuzzy co-cluster containing all documents and words where imposing Ruspini's condition[6] on both document and word membership (in fuzzy co-clustering with Ruspini's condition) make it for word membership in co-cluster must be equal to 1 (in FCCM ,fuzzy codok, fcc_stf).

3. FCCM algorithm have of overflow problem which is mitigating by fuzzy codok but fuzzy codok allowed negative value therefore clipping and renormalization is require in the optimization.

4. In FCCM fuzzifier fuzzy entropy, in fuzzy codok fuzzifier fuzzy gini index , in FCC_STF we replace fuzzy codok's fuzzifirer with a new single term fuzzifier.

5. To reduce outlier fuzzy membership of each web document is found on the priority basic for search engine showing five frame of high priority along with one additional for "zero" priority web page(for outlier).

6. To reduce the problem of curse of dimensionality[5] use FCC_STF, show high performance by experimental result of values (A1,B1,Vcj,A2,B2) seem to less sensitive to local maxima this suggest that hyper dimensionality surface of FCC_STF.

7. The optimization search space is obtained by simplified 2-D plot of J(FCC_STF) and Vcj.

## VI. FUTURE SCOPE

1. Search engine efficiency can be improve by using algorithm FIFO (first in first out), LRU (least recently use) , or the no. of visit for particular web page and frame size can be set accordingly .

2. For search engine we can use standard like XML format which can retrieve page in any language like English, French, italic, Germany etc., this will end communication barrier to retrieve web page.

3. In FCC_STF by using different fuzzifier parameter can improve efficiency of algorithm and also value of objective function along with optimize space for search in 2-D when graph is plotted between Vcj and J(fcc_stf).

4. Even by using web usage mining approach, client visit can be recorded in cookies and on the bases of test case can be made and fuzzy co-clustering can use it to improve retrieval efficiency.

## REFERENCES

[1] J. Han and M. Kamber, Data Mining: Concepts and Techniques Morgar Kauffmann, pp, 21-25, 2000.Proceedings of the Second International Conference on Knowledge Discovery and Data Mining (KDD-96).

[2] Raymond Kosala and Hendrik Blockeel ACM SIGKDD, July 2000 Presented by Shan Huang, 4/24/2007Revised and presented by Fan Min, 4/22/2009.

[3] Web Usage Mining: By BamshadMobasher Dept. of Computer Science, DePaul University, Chicago.

[4] Fuzzy co-clustering of document and key word :Krishna Kummamuru, Ajay Dhawale, and Raghu Krishnapuram ,IBM India Research Lab ,Block 1, IIT, HauzKhas.

[5] Fuzzy Co-clustering of Web Document William-Chandra Tjhi and Lihui Chen Nanyang Technological University, Singapore School of Electrical & Electronic Engineering, Division of Information Engineering.

[6] An introduction to information retrieval (e – book)Robert Cooley, Jaideep Srivastava Dept. of Computer Science, University of Minnesota, Minneapolis, MN cooley@cs.umn.edu, srivasta@cs.umn.edu.

[7] Kumar, S.; Kathuria, M.; Gupta, AK.; Rani, M., "Fuzzy clustering of web documents using equivalence relations and fuzzy hierarchical clustering," Software Engineering (CONSEG), 2012 CSI Sixth International Conference on , vol., no., pp.1,5, 5-7 Sept. 2012.

[8] Rani, Monika, Satendra Kumar, and Vinod Kumar Yadav. "Optimize space search using fcc_stf algorithm in fuzzy co-clustering through search engine."*International Journal of Advanced Research in Computer Engineering & Technology* 1 (2012): 123-127.

[9] Rafael León, J. Javier Rainer, José Manuel Rojo, Ramón Galán: Improving Web Learning through model Optimization using Bootstrap for a Tour-Guide Robot. IJIMAI 1(6): 13-19 (2012).

[10] Juan Carlos Piedra Calderón, J. Javier Rainer: Global Collective Intelligence in Technological Societies: as a result of Collaborative Knowledge in combination with Artificial Intelligence. IJIMAI 2(4): 76-80 (2013).

[11] Juan Carlos Fernandez Rodríguez, J. Javier Rainer, Fernando Miralles Muñoz: Engineering Education through eLearning technology in Spain.IJIMAI 2(1): 46-50 (2013).

[12] Daniel Zapico Palacio, Rubén González Crespo, Gloria García Fernández, Ignacio Rodríguez Novelle: Image/Video Compression with Artificial Neural Networks. IWANN (2) 2009: 330-337.

[13] Daniel Zapico Palacio, Victor Hugo Medina García, Oscar Sanjuán Martínez, Rubén González Crespo: Generating Digital Fingerprints using Artificial Neural Networks. IC-AI 2008: 907-912

[14] Dubey, Shiv Ram, et al. "Infected Fruit Part Detection using K-Means Clustering Segmentation Technique." *IJIMAI* 2.2 (2013): 65-72.

[15] Bhaska, Vinay S., et al. "Business and Social Behaviour Intelligence Analysis Using PSO." *IJIMAI* 2.6 (2014): 69-74.

[16] Bhaskar-Semwal, V., et al. "Accurate location estimation of moving object In Wireless Sensor network." *International Journal of Interactive Multimedia and Artificial Intelligence* 1.4 (2011).

[17] Sati, Meenakshi, et al. "A Fault-Tolerant Mobile Computing Model Based On Scalable Replica." *IJIMAI* 2.6 (2014): 58-68.

[18] SusheelKumar, K., et al. "Generating 3D model using 2D images of an Object."*International Journal of Engineering Science and Technology (IJEST)* 3.1 (2011): 406-415.

[19] Crespo, Rubén González, et al. "Dynamic, ecological, accessible and 3D Virtual Worlds-based Libraries using OpenSim and Sloodle along with mobile location and NFC for checking in." *IJIMAI* 1.7 (2012): 63-69.

[20] Kumar, K. Susheel, et al. "Sports Video Summarization using Priority Curve Algorithm." *International Journal on Computer Science & Engineering* (2010).

[21] Broncano, C. J., et al. "Relative radiometric normalization of multitemporal images." *International Journal of Interactive*

*Multimedia and Artificial Intelligence* 1.3 (2010).

[22] Kumar, K. Susheel, Vijay Bhaskar Semwal, and R. C. Tripathi. "Real time face recognition using adaboost improved fast PCA algorithm." *arXiv preprint arXiv:1108.1353* (2011).

[23] Bijalwan V. et al., (2014), Machine learning approach for text and document mining, arXiv preprint arXiv:1406.1580, 2014(2014).

[24] Vaish, Karan, et al. "Design and Autonomous Control of the Active Adaptive Suspension System of Rudra–Mars Rover." International Journal of Hybrid Information Technology 7.3 (2014).

[25] Gupta J.P. et al., (2013), Human Activity Recognition using Gait Pattern, International Journal of Computer Vision and Image Processing, 3(3), 31 – 53.

[26] Lorenzo, W., Rubén González Crespo, and A. Castillo. "A prototype for linear features generalization." *International Journal of Interactive Multimedia and Artificial Intelligence* 1.3 (2010).

[27] Sanjuan-Martinez, Oscar, et al. "Using Recommendation System for E-learning Environments at degree level." *International Journal of Interactive Multimedia and Artificial Intelligence* 1.2 (2009).

[28] Alexander Pacheco, Holman Bolívar Barón, Rubén González Crespo, Jordán Pascual Espada: Reconstruction of High Resolution 3D Objects from Incomplete Images and 3D Information. IJIMAI 2(6): 7-16 (2014).

[29] Guillermo Cueva-Fernandez, Jordán Pascual Espada, Vicente García-Díaz, Martin Gonzalez-Rodriguez: Kuruma: The Vehicle Automatic Data Capture for Urban Computing Collaborative Systems. IJIMAI 2(2): 28-32 (2013)